\documentclass[a4paper,11pt]{article}
\usepackage{pos}
\usepackage{siunitx}
\usepackage{caption}
\usepackage{subcaption}
\usepackage{gensymb}
\usepackage{wrapfig}
\usepackage{todonotes}
\usepackage{enumitem}
\usepackage{lineno}
\usepackage{graphicx}\graphicspath{{figures/}}
\usepackage[capitalize]{cleveref}

\title{Observation of the Shadows of the Moon and Sun Using the Pierre Auger Observatory at an Average Energy of $7\times10^{17}\,$eV}
\ShortTitle{The Shadows of the Moon and Sun with the Pierre Auger Observatory}

\author*[ab]{Katar\'ina~\v Simkov\'a}

\affiliation[a]{Vrije Universiteit Brussel, Physics Department, Pleinlaan 2, 1050 Brussels, Belgium}
\affiliation[b]{Universit\'e Libre de Bruxelles, Physics Department, Boulevard du Triomphe 2, 1050 Brussels, Belgium }

\onbehalf{for the Pierre Auger Collaboration$^c$}
\affiliation[c]{Observatorio Pierre Auger, Av.\ San Mart{\'\i}n Norte 304, 5613 Malarg\"ue, Argentina\\
Full author list: {\rm\url{https://www.auger.org/archive/authors_icrc_2025.html}}}

\emailAdd{spokespersons@auger.org}

\abstract{The interaction of cosmic rays with celestial bodies such as the Moon or
the Sun produces a shadow in the arrival direction distribution of the
cosmic rays reaching the Earth. Such deficits from an isotropic flux
have been observed by astroparticle observatories below energies of
$10^{15}\,$eV. Above this energy, measurements were limited due to the
low number of events as a result of the steeply falling cosmic-ray flux
with energy. With more than 10.6 million events recorded during 20 years
of operation of the Pierre Auger Observatory, we report the first
observation of the shadow of the Moon at an average energy of
$7\times10^{17}\,$eV with a maximum significance above 3$\sigma$.
The shadow is an end-to-end check that the celestial directions are correctly reconstructed from the air shower data, and it is used here to derive the effective angular resolution for this dataset. Additionally, we present the results of a similar study on the
shadow of the Sun.}

\ConferenceLogo{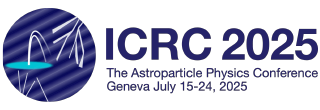}

\FullConference{39th International Cosmic Ray Conference (ICRC2025)\\
 15–24 July 2025\\
Geneva, Switzerland\\}

\begin{document}
\maketitle

\section{Introduction}
The Moon and the Sun, as close-by celestial objects, remove cosmic rays from the approximately isotropic flux, creating shadows in the flux seen from Earth~\cite{Clark57}. With a perfect observatory resolution, these shadows appear as disks with an angular radius of the Moon and the Sun, both approximately \qty{0.26}{\degree} on average. At TeV-PeV energies, the shape and location of the shadows are further distorted by deviations in the magnetic fields of the Sun and Earth, while at higher energies these effects are expected to be negligible~\cite{CRpropagSun,Aartsen_2014_icecube}. The final shape of the observed shadows is then influenced by the pointing and resolution of an observatory. That is why the shadows of the Moon and the Sun observed in TeV-PeV cosmic rays have been a commonly used technique to verify the performance of astroparticle observatories~\cite{Aartsen_2014_icecube,PhysRevD.49.1171_CASA,PhysRevD.43.1735_Cygnus,1993ICRC....4..351A_Tibet,Achard_2005,Nan:2021smu_LHAASO}. Because the flux of cosmic rays drops steeply with increasing energy, the observation above \qty{e16}{\electronvolt} has not been possible until now. Using the large exposure of the Pierre Auger Observatory, we report the first observation of the shadows at an average energy of $7\times10^{17}\,$eV. In the last part of this proceeding, we use the shape of the deficit to evaluate the effective angular resolution as a combination of resolutions of events close to the Moon and Sun.

\section{The Pierre Auger Observatory and the Data}
The Pierre Auger Observatory, located in the southern hemisphere close to the city of Malarg\"ue, in Argentina's Mendoza province, has been measuring extensive air showers (EASs) since 2004. It is composed of two key detectors. The first is the surface detector (SD) with a duty cycle of $\sim$\qty{100}{\percent} measuring the particles reaching the ground. The second is the fluorescence detector (FD), which measures the development of EAS in the atmosphere and provides an almost calorimetric energy estimate. The energy scale of SD is cross-calibrated with the FD measurements to obtain the energy for events measured by SD only \cite{2015172_NIM_Auger}. 

\begin{figure}[!h]
    \centering
        \includegraphics[width=\textwidth]{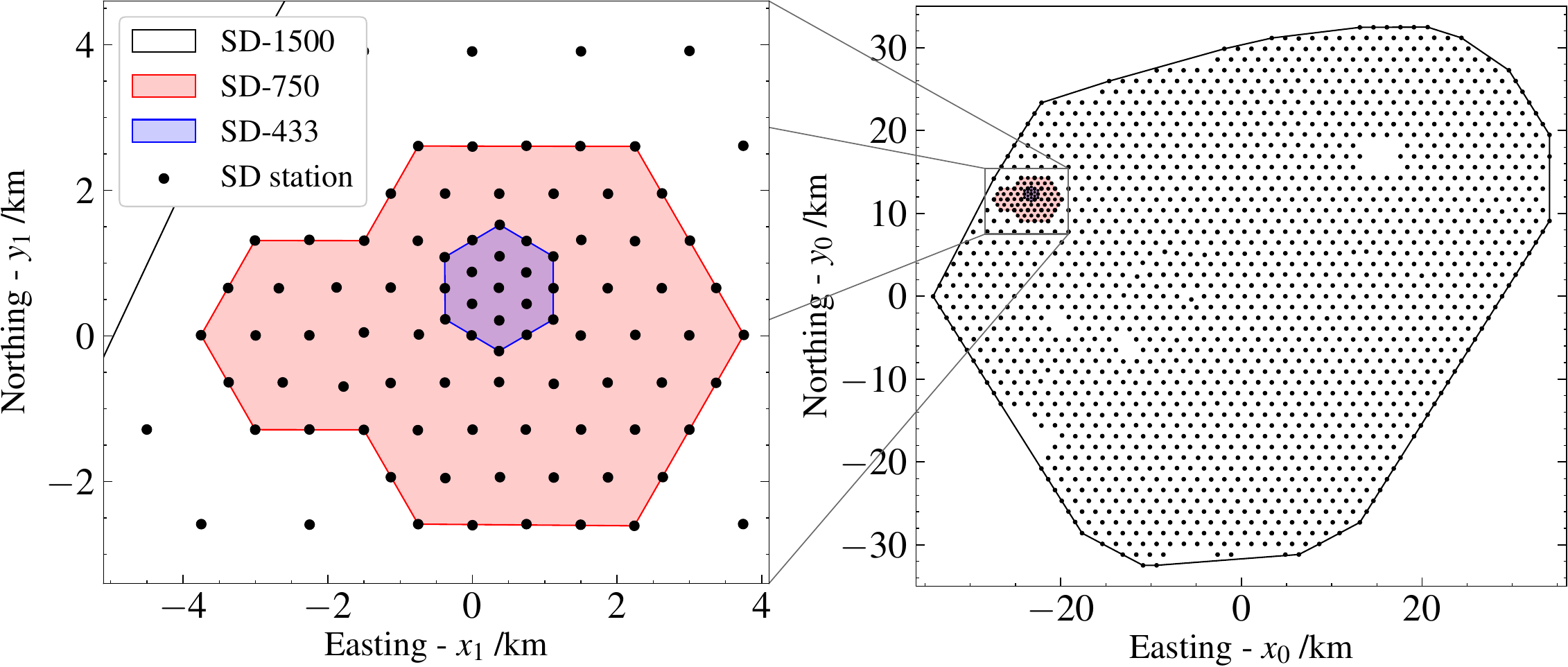}
      \caption{The layout of the SD-750 and SD-433 (red and blue region in the left panel) placed within the SD-1500 array (bounded by the black line in the right panel). The coordinates are centered at $(x_1,y_1)=(450627, 6113915)$\,m and $(x_0,y_0)=(474255, 6102240)$\,m.}
      \label{fig:arrays_plot}
    \end{figure}

The data used in this study come from the SD. Its main particle detector is a cylindrical water-Cherenkov detector (WCD) of \qty{1.2}{\meter} height and \qty{3.6}{\meter} diameter overlooked by three photomultipliers. During a recently-finished upgrade, the SD stations were further equipped by a Surface-Scintillator Detector (SSD) and a Radio Detector (RD), both placed on top of the WCD. The data in this study come from before the upgrade and consider signals from the WCDs only. The stations are placed on a triangular grid of regular spacing. Three arrays are defined according to the distances between the stations and their layout is shown in \cref{fig:arrays_plot}. The largest one, SD-1500, consists of \num{1600} WCDs with a spacing of \qty{1500}{m} that span an area of \qty{3000}{\square\kilo\meter} measuring ultra-high-energy cosmic rays down to \qty{e18}{\electronvolt}. The SD-750 is located within the SD-1500. It is composed of \num{61} stations with a spacing of \qty{750}{\meter} and covers area of \qty{27}{\square\kilo\meter}. It is most sensitive to measure cosmic-ray primaries in the energy range between \qty{e17}{\electronvolt} and \qty{e18}{\electronvolt}. The last WCD array is SD-433 with \qty{433}{m} spacing between \num{19} stations on \qty{2}{\square\kilo\meter} placed within SD-750. The SD-433 extends the sensitivity of the Observatory down to \qty{e16}{\electronvolt} \cite{Silli:2021Jt_SD-433}.

To obtain high-level variables such as arrival direction and the energy of a primary cosmic ray, the reconstruction of events is based on the same principle in all three arrays. At least three triggered WCDs are required to reconstruct the arrival direction based on the signal start time in each detector. The more WCDs that are used in the reconstruction, the more accurate is the reconstructed arrival direction. Previous studies, based on both measurements and simulations of SD-1500, estimated the angular resolution to be around \qty{1}{\degree}, improving to \qty{0.4}{\degree} with higher number of stations and energy~\cite{Aab_2020_Auger_reco}.

In this study, we use \num{10.6} million events recorded by all three SD arrays between January 2004 and March 2023\footnote{The time range is from the start of data-taking of each array (SD-1500: 2004, SD-750: 2008, SD-433: 2018) until the upgrade of the Observatory (AugerPrime).}. The biggest contribution comes from SD-750 (\qty{64}{\percent} of the data), followed by SD-1500 (\qty{29}{\percent}) and SD-433 (\qty{7}{\percent}). Further, we select events that fall within \qty{5}{\degree} great-circle angular distance from the center of the Moon/Sun in local coordinates using the arrival direction of the event from the reconstruction and the position of the Moon/Sun from \texttt{Astropy} Python library~\cite{astropy:2022}. This results in \num{18215} events for the study of the lunar shadow and \num{18650} events for the solar shadow.

\section{The Relative Deficits due to the Moon and the Sun and their Significance}

\begin{figure}[!th]
    \centering
        \includegraphics[width=0.7\textwidth]{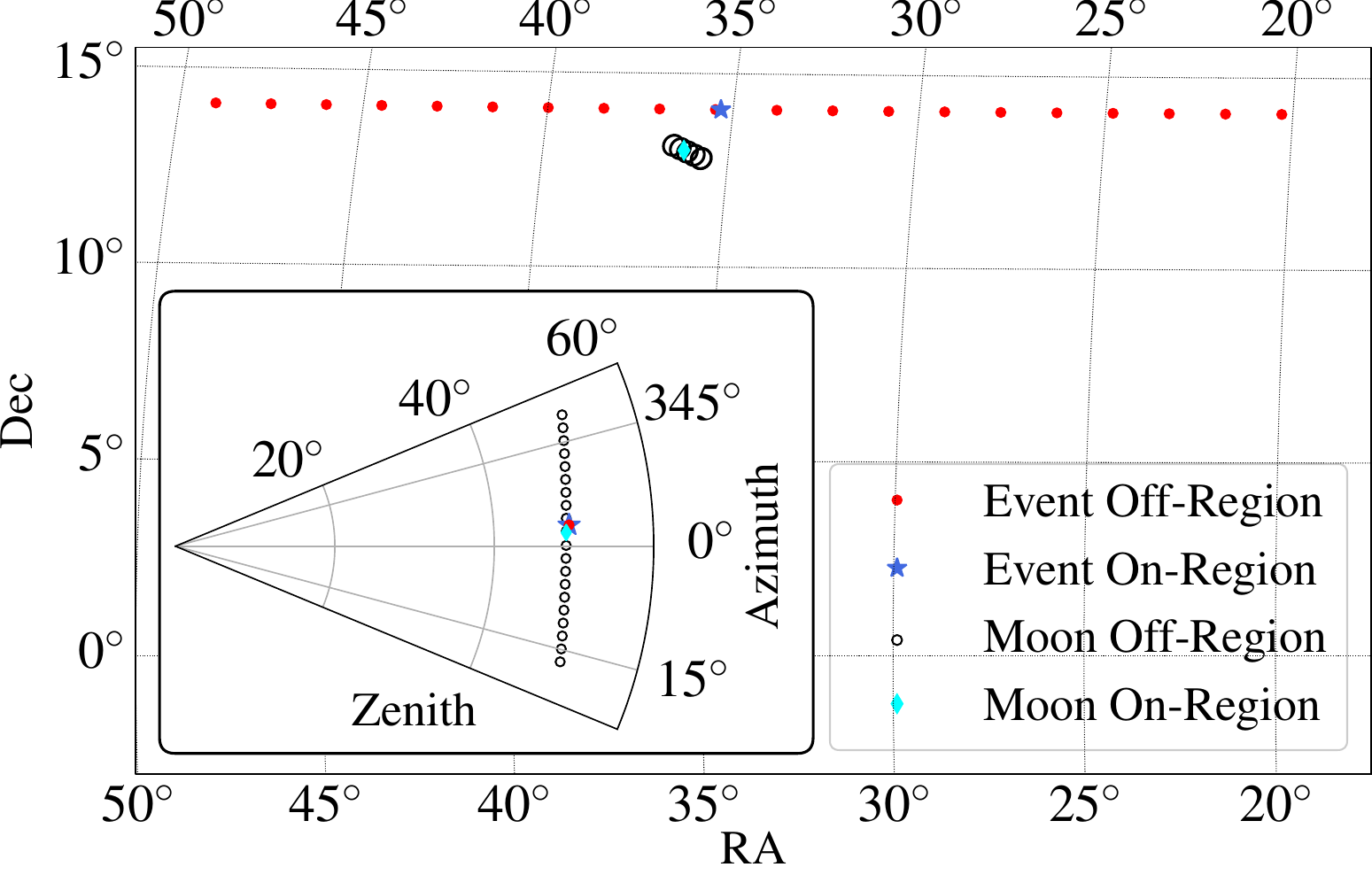}
      \caption{Example of the coordinate changes using the time shuffling method to generate the isotropic background estimate (average off-region), only every tenth off-region is shown. While the celestial object has approximately the same position in the equatorial coordinates, the off-events move in RA keeping the same declination which ensures the correct exposure estimate in the off-regions. In local coordinates, the position of the event is preserved while the fake moons move across the sky.}
      \label{fig:method}
    \end{figure}

The flux in the direction of the Moon and Sun (on-region) is compared to an isotropic flux expectation. The isotropic expectation is deduced from generated sky regions that contain fake moons or suns (off-regions). Due to the fixed location of the Observatory and the 100\% duty cycle of the SD, the exposure depends just on the declination (Dec) and it is uniform in right ascension (RA).
We generate the off-regions using a time-shuffling method. By doing so, the time of an event is changed evenly 200 (100) times in $\pm1\,$h ($\pm30\,$min) for the Moon (Sun)\footnote{Using larger time interval for the Sun biases the off-region. Events close to the Sun are detected only during the day, which may introduce a dependence on RA.}. The position of the events is changed in RA while its Dec remains the same. In local coordinates (Alt, Az) by changing the time, the Moon and the Sun move across the sky while for the event, Alt \& Az are preserved. An example of the change of coordinates is shown in \cref{fig:method}. With $s$ off-regions (200 for the Moon and 100 for the Sun), the reference expectation is obtained as the average of the off-regions. 

\begin{figure}[!th]
\centering
\begin{subfigure}{0.49\linewidth}
    \centering
    \includegraphics[width=\textwidth,page=1]{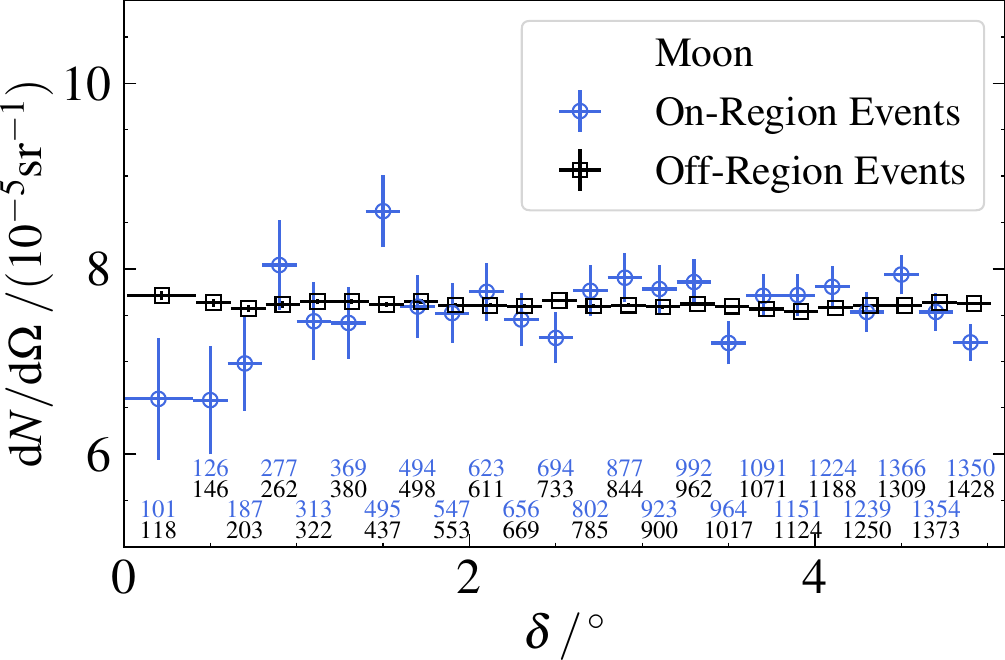}
\end{subfigure}
\begin{subfigure}{0.49\linewidth}
    \centering
    \includegraphics[width=\textwidth,page=2]{dNdOmega_moon_sun.pdf_2p}
\end{subfigure} 
\caption{Binned numbers of events in annuli centered on the Moon (left) and Sun (right) normalized by the solid angles. The off-region numbers are averaged over the off-regions. The absolute numbers of events in each bin are given with on-region event number in color and the averaged off-region number in black.}
\label{fig:dNdOmega}
\end{figure}

The numbers of events in the on- and off-region binned according to the angular distance ($\delta$) between the center of the Moon/Sun and the events are shown in~\cref{fig:dNdOmega}. The numbers are normalized by the solid angle. The uncertainties of the numbers go as square-root of the number for the on-region, while for the off-region the uncertainty is further divided by the square-root of the number of off-regions $s$. For the off-region, the event number per solid angle shows no dependency on the angular distance as expected. At larger angular distances, the on-region numbers are consistent with the off-region. This confirms that our off-region estimate is unbiased. Close to the Moon, a deficit appears in the first three bins, i.e. up to \qty{0.8}{\degree}. For the Sun the deficit appears in the first bin ($\delta<0.4\degree$) and the third bin ($0.6\degree<\delta<0.8\degree$).

To better quantify the deficits, we use the relative difference of events compared to the off-region and we sum the number of events below a certain angular distance $\delta$. The relative difference is given by:
\begin{equation}
    \Delta(\delta) = \frac{N_\text{on}(\delta)}{N_\text{off}(\delta)} -1 \pm
    \frac{N_\text{on}(\delta)}{N_\text{off}(\delta)}
    \sqrt{\frac{1}{N_\text{on}(\delta)} + \frac{1}{s~N_\text{off}(\delta)}}\,.
\end{equation}

\begin{figure}[!th]
\centering
\begin{subfigure}{0.49\linewidth}
    \centering
    \includegraphics[width=\textwidth,page=1]{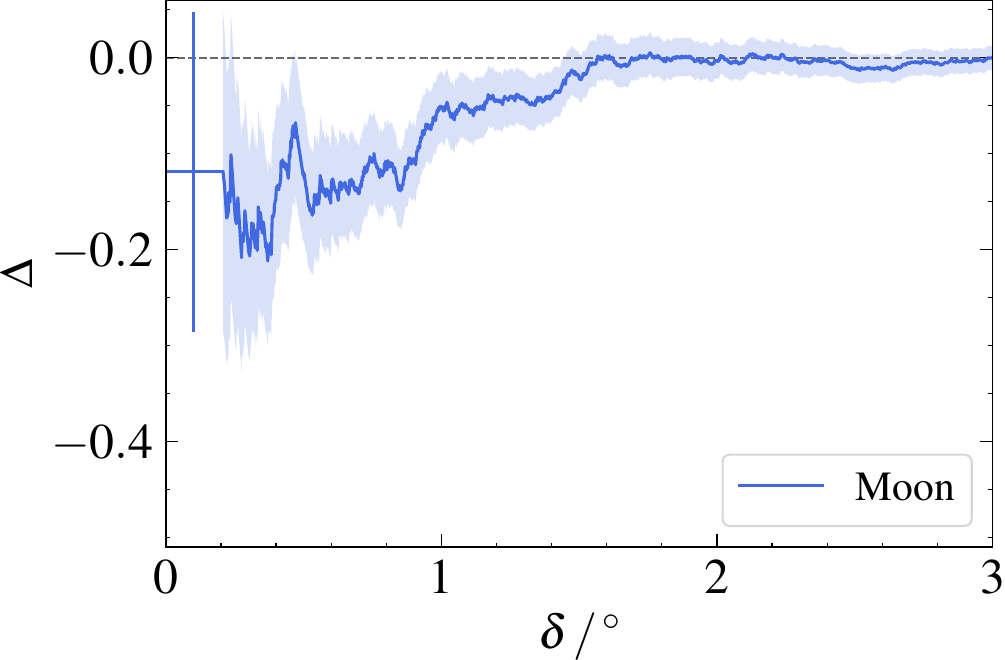}
\end{subfigure}
\begin{subfigure}{0.49\linewidth}
    \centering
    \includegraphics[width=\textwidth,page=2]{cumul_moon_sun_2p}
\end{subfigure} 
\caption{The relative difference of the cumulative number of events in the `on' and `off' regions as a function of the angular distance to the Moon (left panel) and to the Sun (right panel). Shaded bands indicate the $1\sigma$ statistical uncertainty. With the relative difference being cumulative, the individual data points are correlated.}
\label{fig:cumul_moon_sun}
\end{figure}

\begin{figure}[!th]
\centering
\includegraphics[width=0.5\columnwidth]{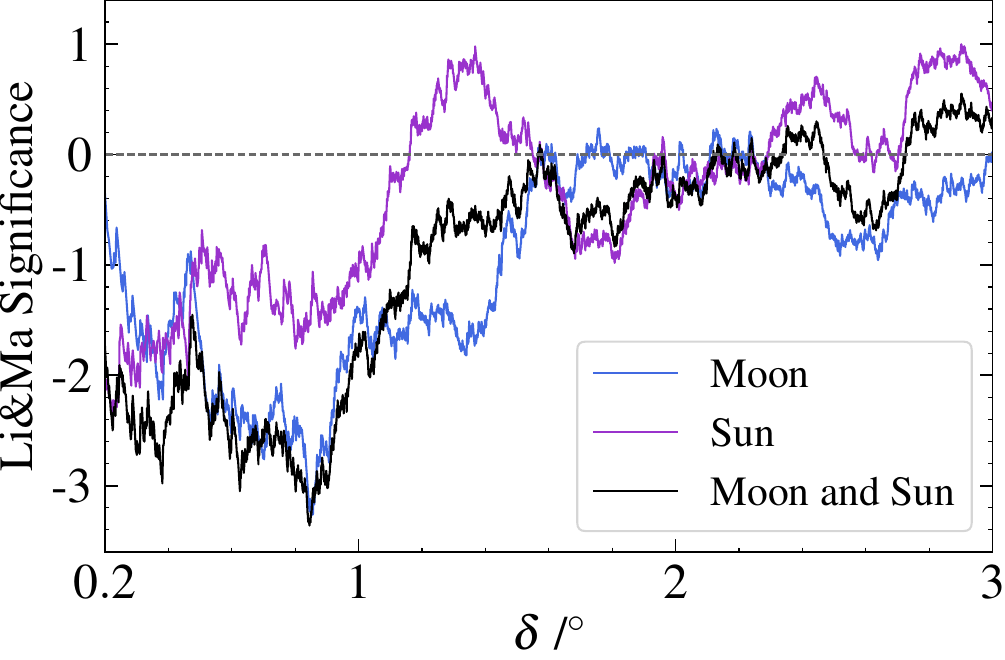}
\caption{Individual and combined Li\&Ma significances of the number of events close to the Moon/Sun compared to the isotropic expectation as a function of the angular distance between the events and the centers of the Moon/Sun.}
\label{fig:signif_moon_sun}
\end{figure}

These cumulative relative differences for the Moon and the Sun are shown in \cref{fig:cumul_moon_sun}. Both exhibit a deficit below \qty{1.4}{\degree} confirming accurate pointing of the Observatory. The significance of the shadows is first calculated using the Li\&Ma significance \cite{LiMa} and it is shown in \cref{fig:signif_moon_sun}. The maximum significance appears at $\delta=0.85\degree$ for both the Moon and the combined data, with values of $3.28\sigma$ and $3.36\sigma$ respectively. At this $\delta$, the shadow of the Sun with the maximum significance of $2.41\sigma$ at $\delta=0.23\degree$, contributes only little to the combined significance.

\section{The Effective Angular Resolution of the Observatory}
\begin{figure}[!th]
\centering
\begin{subfigure}{0.49\linewidth}
    \centering
    \includegraphics[width=\textwidth,page=1]{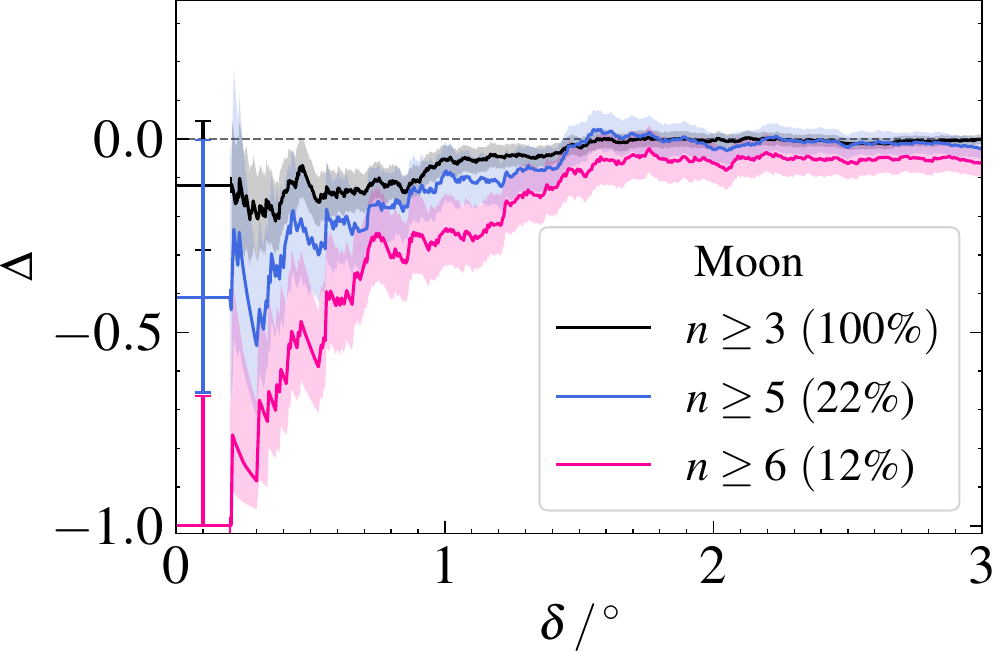}
\end{subfigure}
\begin{subfigure}{0.49\linewidth}
    \centering
    \includegraphics[width=\textwidth,page=2]{cumul_multip_2p}
\end{subfigure} 
\caption{The relative difference of the cumulative number of events as a function of angular distance for events with different detector multiplicities $n$; left panel: Moon, right panel: Sun.
The fractions of the data samples compared to the full data set are quoted in the legend.}
\label{fig:cumul_Nstation}
\end{figure}

As described in the introduction, the shape of the deficit is influenced by the resolution of the Observatory. The better the resolution, the fewer detected events migrate inside the shadow. Because the resolution improves with the multiplicity of stations participating in an event $n$, we divided the data set in samples based on the multiplicity. The deficit for these subsamples is illustrated in \cref{fig:cumul_Nstation}. We observe a decrease of $\Delta$ with increasing multiplicity, as expected when the angular resolution is improving. Due to the low statistics, the ordering is not very significant: At \qty{0.5}{\degree}, the deficit due to the Moon grows from $(12\pm7)$\% for all the data, to $(26\pm13)$\% with $n\geq5$ and $(54\pm15)$\% with $n\geq6$. For the Sun the deficits at \qty{0.5}{\degree} are $(6\pm7)$\% for all the data, $(28\pm13)$\% for $n\geq5$ and $(3\pm22)$\% for $n\geq6$.

Next, we deduce the effective angular resolution of the Observatory based on the unbinned fit of the distribution of events around the center of the Moon and the Sun. The likelihood is constructed as in \cite{PhysRevD.49.1171_CASA}, where the shadowed events, dispersed by a 2-D Gaussian point spread function (PSF) are subtracted from the isotropic flux constant in $\cos\delta$. The number of events $N$ in an infinitesimal annulus $\mathrm{d}\cos{\delta}$ is given by:
\begin{equation}\label{eq:likelihood}
    \frac{\mathrm{d}N}{\mathrm{d}\cos\delta} =
  1 -
  \int^{\delta_\text{c}}_0 \!\!\!\!\! r'\,\mathrm{d}r'
    \int^{2\pi}_0 \!\!\! \mathrm{d}\phi' \, \frac{1}{2\pi\sigma_\text{r}^2}\exp\left[-\frac{\xi(\cos\delta;r',\phi')^2}{2\sigma_\text{r}^2}\right].
\end{equation}
The integration goes over the disk of the Moon/Sun in polar coordinates $(r',\phi')$ with $r'$ ranging from zero up to their angular radius $\delta_\text{c}=\ang{0.26}$. $\xi(\cos\delta;r',\phi')$ is the angular distance between an event and the infinitesimal element of the disk. The probability $P$ is normalized to take into account that we consider only events up to $\delta=5\degree$:
\begin{equation}
    P =
    \frac{\mathrm{d}P/\mathrm{d}\cos\delta}{\int^{\cos(5\degree)}_1(\mathrm{d}P/\mathrm{d}\cos\delta) \, \mathrm{d}\cos\delta}
\quad\text{with}\quad
  \frac{\mathrm{d}P}{\mathrm{d}\cos\delta} =
    \frac{\mathrm{d}N}{\mathrm{d}\cos\delta}
    \frac{\int^{\cos(5\degree)}_1\mathrm{d}\cos\delta}{\int^{\cos(5\degree)}_1(\mathrm{d}N/\mathrm{d}\cos\delta) \,
    \mathrm{d}\cos\delta}.
\end{equation}
The only free parameter in the fit is $\sigma_\mathrm{r}$. We define the effective angular resolution of the Observatory as $\sigma_{68}=\sqrt{2.278}\sigma_\mathrm{r}$ containing 68\% probability of the 2-D Gaussian PSF. The result of the fit is $\sigma_{68}=(0.61^{+0.21}_{-0.13})\degree$ for the Moon data, $\sigma_{68}=(0.54^{+0.30}_{-0.15})\degree$ for the Sun and $\sigma_{68}=(0.59^{+0.15}_{-0.11})\degree$ for the combined data. The uncertainties are obtained from the shape of the likelihood and were further verified using a bootstrap of the data.

The significance of the fits ($N_\sigma$) can be evaluated using the likelihood ratio test as in \cite{PhysRevD.43.1735_Cygnus} with no shadows (i.e. a very poor resolution) being the null hypothesis:
\begin{equation}
    N_\sigma = \sqrt{2(w_\mathrm{max}-w_\infty)}\,.
\end{equation}
 Here $w_\mathrm{max}$ is the log-likelihood of the best parameter value and $w_\infty$ the log-likelihood of a poor resolution, we used $\sigma_\mathrm{r}=10\degree$. The resulting significance of the fit of the Moon data $N_\sigma=2.4$, for the Sun it is $N_\sigma=1.8$, while for the combined data $N_\sigma=3.0$.

The effective angular resolution represents the combination of multiple resolutions inherent to different multiplicities, zenith angles and the three arrays. Therefore, we performed a toy Monte-Carlo study to estimate the expected effective angular resolution based on the event-level resolutions assigned during the event reconstruction. The most probable value of the effective angular resolution of $(0.
65\pm0.02)\degree$ is consistent with all the three fit results, validating the event-level resolution estimates.

According to the $\sigma_{68}$ of the fit of the shadow of the Moon, the maximum significance is expected at $\delta=(0.68^{+0.22}_{-0.14})\degree$. This is consistent with the observed maximum of the Li\&Ma significance at \qty{0.85}{\degree} (\cref{fig:signif_moon_sun}). 

\section{Conclusion}

Analyzing more than 10.6 million events of the Pierre Auger Observatory above \qty{e16}{\electronvolt}, we observe for the first time the shadow of the Moon and the Sun in such high-energy cosmic rays. The significance of the combined likelihood fit reaches $3\sigma$. The contribution from the deficit of the Moon is larger, with a maximum Li\&Ma significance of $3.28\sigma$ at \qty{0.85}{\degree} angular distance. The appearance of the deficit in our data validates the accurate pointing of the Observatory. 

Based on the size of the shadow of the Moon, i.e. the likelihood fit, we find an effective angular resolution of $\sigma_{68}=(0.61^{+0.21}_{-0.13})\degree$ for our Observatory. It represents a combination of resolutions varying in multiplicity, zenith angle and energy. With this resolution, the maximum significance of the shadows is expected at $\delta=(0.68^{+0.22}_{-0.14})\degree$, which is consistent with our observation of the maximum Li\&Ma significance at $\delta=0.85\degree$.

\bibliographystyle{JHEP}
\bibliography{my_bib}

\providecommand{\href}[2]{#2}\begingroup\raggedright\begin{thebibliography}{10}

\bibitem{Clark57}
G.W.~Clark, \emph{Arrival directions of cosmic-ray air showers from the northern sky}, \href{https://doi.org/10.1103/PhysRev.108.450}{\emph{Phys. Rev.} {\bfseries 108} (1957) 450}.

\bibitem{CRpropagSun}
{Becker Tjus, J.}, {Desiati, P.}, {Döpper, N.}, {Fichtner, H.}, {Kleimann, J.}, {Kroll, M.} et~al., \emph{Cosmic-ray propagation around the {Sun}: investigating the influence of the solar magnetic field on the cosmic-ray {Sun} shadow}, \href{https://doi.org/10.1051/0004-6361/201936306}{\emph{A\&A} {\bfseries 633} (2020) A83}.

\bibitem{Aartsen_2014_icecube}
{\scshape IceCube} collaboration, \emph{Observation of the cosmic-ray shadow of the {Moon} with {IceCube}}, \href{https://doi.org/10.1103/physrevd.89.102004}{\emph{Physical Review D} {\bfseries 89} (2014) }.

\bibitem{PhysRevD.49.1171_CASA}
{\scshape The Chicago Air Shower Array} collaboration, \emph{Observation of the shadows of the {Moon} and {Sun} using 100 {TeV} cosmic rays}, \href{https://doi.org/10.1103/PhysRevD.49.1171}{\emph{Phys. Rev. D} {\bfseries 49} (1994) 1171}.

\bibitem{PhysRevD.43.1735_Cygnus}
D.E.~Alexandreas et~al., \emph{Observation of shadowing of ultrahigh-energy cosmic rays by the {Moon} and the {Sun}}, \href{https://doi.org/10.1103/PhysRevD.43.1735}{\emph{Phys. Rev. D} {\bfseries 43} (1991) 1735}.

\bibitem{1993ICRC....4..351A_Tibet}
M.~{Amenomori} et~al., \emph{{Cosmic Ray Shadow by the {Moon} Observed with the {Tibet Air Shower Array}}},  in \emph{23rd International Cosmic Ray Conference (ICRC23), Volume 4}, D.A.~{Leahy}, R.B.~{Hicks} and D.~{Venkatesan}, eds., vol.~4 of \emph{International Cosmic Ray Conference}, p.~351, Jan., 1993.

\bibitem{Achard_2005}
P.~Achard, O.~Adriani, M.~Aguilar-Benitez, M.~van~den Akker, J.~Alcaraz, G.~Alemanni et~al., \emph{Measurement of the shadowing of high-energy cosmic rays by the moon: A search for tev-energy antiprotons}, \href{https://doi.org/10.1016/j.astropartphys.2005.02.002}{\emph{Astroparticle Physics} {\bfseries 23} (2005) 411–434}.

\bibitem{Nan:2021smu_LHAASO}
{\scshape LHAASO} collaboration, \emph{{The performances of the {LHAASO-KM2A} tested by the observation of cosmic-ray {Moon} shadow}}, \href{https://doi.org/10.22323/1.395.0350}{\emph{PoS} {\bfseries ICRC2021} (2021) 350}.

\bibitem{2015172_NIM_Auger}
{\scshape The Pierre Auger} collaboration, \emph{{The Pierre Auger Cosmic Ray Observatory}}, \href{https://doi.org/https://doi.org/10.1016/j.nima.2015.06.058}{\emph{Nuclear Instruments and Methods in Physics Research Section A: Accelerators, Spectrometers, Detectors and Associated Equipment} {\bfseries 798} (2015) 172}.

\bibitem{Silli:2021Jt_SD-433}
{\scshape The Pierre Auger} collaboration, \emph{{Performance of the 433 m surface array of the {Pierre Auger Observatory}}}, \href{https://doi.org/10.22323/1.395.0224}{\emph{PoS} {\bfseries ICRC2021} (2021) 224}.

\bibitem{Aab_2020_Auger_reco}
{\scshape The Pierre Auger} collaboration, \emph{Reconstruction of events recorded with the surface detector of the {Pierre Auger Observatory}}, \href{https://doi.org/10.1088/1748-0221/15/10/P10021}{\emph{Journal of Instrumentation} {\bfseries 15} (2020) P10021}.

\bibitem{astropy:2022}
{Astropy Collaboration} and {Astropy Project Contributors}, \emph{{The Astropy Project: Sustaining and Growing a Community-oriented Open-source Project and the Latest Major Release (v5.0) of the Core Package}}, \href{https://doi.org/10.3847/1538-4357/ac7c74}{\emph{Astrophysical Journal} {\bfseries 935} (2022) 167} [\href{https://arxiv.org/abs/2206.14220}{{\ttfamily 2206.14220}}].

\bibitem{LiMa}
T.P.~{Li} and Y.Q.~{Ma}, \emph{{Analysis methods for results in gamma-ray astronomy.}}, \href{https://doi.org/10.1086/161295}{\emph{Astrophysical Journal} {\bfseries 272} (1983) 317}.

\end{thebibliography}\endgroup

\newpage
\section*{The Pierre Auger Collaboration}

{\footnotesize\setlength{\baselineskip}{10pt}
\noindent
\begin{wrapfigure}[11]{l}{0.12\linewidth}
\vspace{-4pt}
\includegraphics[width=0.98\linewidth]{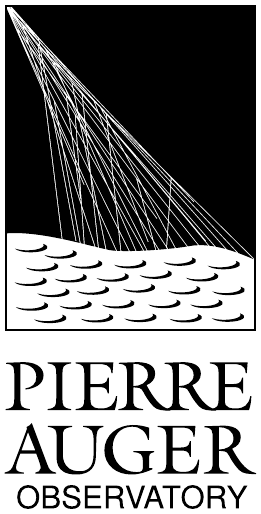}
\end{wrapfigure}
\begin{sloppypar}\noindent
A.~Abdul Halim$^{13}$,
P.~Abreu$^{70}$,
M.~Aglietta$^{53,51}$,
I.~Allekotte$^{1}$,
K.~Almeida Cheminant$^{78,77}$,
A.~Almela$^{7,12}$,
R.~Aloisio$^{44,45}$,
J.~Alvarez-Mu\~niz$^{76}$,
A.~Ambrosone$^{44}$,
J.~Ammerman Yebra$^{76}$
% ...

\end{sloppypar}
\begin{center}
\end{center}

\vspace{1ex}
\begin{description}[labelsep=0.2em,align=right,labelwidth=0.7em,labelindent=0em,leftmargin=2em,noitemsep,before={\renewcommand\makelabel[1]{##1 }}]
\item[$^{1}$] Centro At\'omico Bariloche and Instituto Balseiro (CNEA-UNCuyo-CONICET), San Carlos de Bariloche, Argentina
\item[$^{2}$] Departamento de F\'\i{}sica and Departamento de Ciencias de la Atm\'osfera y los Oc\'eanos, FCEyN, Universidad de Buenos Aires and CONICET, Buenos Aires, Argentina
\item[$^{3}$] IFLP, Universidad Nacional de La Plata and CONICET, La Plata, Argentina
\item[$^{4}$] Instituto de Astronom\'\i{}a y F\'\i{}sica del Espacio (IAFE, CONICET-UBA), Buenos Aires, Argentina
\item[$^{5}$] Instituto de F\'\i{}sica de Rosario (IFIR) -- CONICET/U.N.R.\ and Facultad de Ciencias Bioqu\'\i{}micas y Farmac\'euticas U.N.R., Rosario, Argentina
\item[$^{6}$] Instituto de Tecnolog\'\i{}as en Detecci\'on y Astropart\'\i{}culas (CNEA, CONICET, UNSAM), and Universidad Tecnol\'ogica Nacional -- Facultad Regional Mendoza (CONICET/CNEA), Mendoza, Argentina
\item[$^{7}$] Instituto de Tecnolog\'\i{}as en Detecci\'on y Astropart\'\i{}culas (CNEA, CONICET, UNSAM), Buenos Aires, Argentina
\item[$^{8}$] International Center of Advanced Studies and Instituto de Ciencias F\'\i{}sicas, ECyT-UNSAM and CONICET, Campus Miguelete -- San Mart\'\i{}n, Buenos Aires, Argentina
\item[$^{9}$] Laboratorio Atm\'osfera -- Departamento de Investigaciones en L\'aseres y sus Aplicaciones -- UNIDEF (CITEDEF-CONICET), Argentina
\item[$^{10}$] Observatorio Pierre Auger, Malarg\"ue, Argentina
\item[$^{11}$] Observatorio Pierre Auger and Comisi\'on Nacional de Energ\'\i{}a At\'omica, Malarg\"ue, Argentina
\item[$^{12}$] Universidad Tecnol\'ogica Nacional -- Facultad Regional Buenos Aires, Buenos Aires, Argentina
\item[$^{13}$] University of Adelaide, Adelaide, S.A., Australia
\item[$^{14}$] Universit\'e Libre de Bruxelles (ULB), Brussels, Belgium
\item[$^{15}$] Vrije Universiteit Brussels, Brussels, Belgium
\item[$^{16}$] Centro Brasileiro de Pesquisas Fisicas, Rio de Janeiro, RJ, Brazil
\item[$^{17}$] Centro Federal de Educa\c{c}\~ao Tecnol\'ogica Celso Suckow da Fonseca, Petropolis, Brazil
\item[$^{18}$] Instituto Federal de Educa\c{c}\~ao, Ci\^encia e Tecnologia do Rio de Janeiro (IFRJ), Brazil
\item[$^{19}$] Universidade de S\~ao Paulo, Escola de Engenharia de Lorena, Lorena, SP, Brazil
\item[$^{20}$] Universidade de S\~ao Paulo, Instituto de F\'\i{}sica de S\~ao Carlos, S\~ao Carlos, SP, Brazil
\item[$^{21}$] Universidade de S\~ao Paulo, Instituto de F\'\i{}sica, S\~ao Paulo, SP, Brazil
\item[$^{22}$] Universidade Estadual de Campinas (UNICAMP), IFGW, Campinas, SP, Brazil
\item[$^{23}$] Universidade Estadual de Feira de Santana, Feira de Santana, Brazil
\item[$^{24}$] Universidade Federal de Campina Grande, Centro de Ciencias e Tecnologia, Campina Grande, Brazil
\item[$^{25}$] Universidade Federal do ABC, Santo Andr\'e, SP, Brazil
\item[$^{26}$] Universidade Federal do Paran\'a, Setor Palotina, Palotina, Brazil
\item[$^{27}$] Universidade Federal do Rio de Janeiro, Instituto de F\'\i{}sica, Rio de Janeiro, RJ, Brazil
\item[$^{28}$] Universidad de Medell\'\i{}n, Medell\'\i{}n, Colombia
\item[$^{29}$] Universidad Industrial de Santander, Bucaramanga, Colombia
\item[$^{30}$] Charles University, Faculty of Mathematics and Physics, Institute of Particle and Nuclear Physics, Prague, Czech Republic
\item[$^{31}$] Institute of Physics of the Czech Academy of Sciences, Prague, Czech Republic
\item[$^{32}$] Palacky University, Olomouc, Czech Republic
\item[$^{33}$] CNRS/IN2P3, IJCLab, Universit\'e Paris-Saclay, Orsay, France
\item[$^{34}$] Laboratoire de Physique Nucl\'eaire et de Hautes Energies (LPNHE), Sorbonne Universit\'e, Universit\'e de Paris, CNRS-IN2P3, Paris, France
\item[$^{35}$] Univ.\ Grenoble Alpes, CNRS, Grenoble Institute of Engineering Univ.\ Grenoble Alpes, LPSC-IN2P3, 38000 Grenoble, France
\item[$^{36}$] Universit\'e Paris-Saclay, CNRS/IN2P3, IJCLab, Orsay, France
\item[$^{37}$] Bergische Universit\"at Wuppertal, Department of Physics, Wuppertal, Germany
\item[$^{38}$] Karlsruhe Institute of Technology (KIT), Institute for Experimental Particle Physics, Karlsruhe, Germany
\item[$^{39}$] Karlsruhe Institute of Technology (KIT), Institut f\"ur Prozessdatenverarbeitung und Elektronik, Karlsruhe, Germany
\item[$^{40}$] Karlsruhe Institute of Technology (KIT), Institute for Astroparticle Physics, Karlsruhe, Germany
\item[$^{41}$] RWTH Aachen University, III.\ Physikalisches Institut A, Aachen, Germany
\item[$^{42}$] Universit\"at Hamburg, II.\ Institut f\"ur Theoretische Physik, Hamburg, Germany
\item[$^{43}$] Universit\"at Siegen, Department Physik -- Experimentelle Teilchenphysik, Siegen, Germany
\item[$^{44}$] Gran Sasso Science Institute, L'Aquila, Italy
\item[$^{45}$] INFN Laboratori Nazionali del Gran Sasso, Assergi (L'Aquila), Italy
\item[$^{46}$] INFN, Sezione di Catania, Catania, Italy
\item[$^{47}$] INFN, Sezione di Lecce, Lecce, Italy
\item[$^{48}$] INFN, Sezione di Milano, Milano, Italy
\item[$^{49}$] INFN, Sezione di Napoli, Napoli, Italy
\item[$^{50}$] INFN, Sezione di Roma ``Tor Vergata'', Roma, Italy
\item[$^{51}$] INFN, Sezione di Torino, Torino, Italy
\item[$^{52}$] Istituto di Astrofisica Spaziale e Fisica Cosmica di Palermo (INAF), Palermo, Italy
\item[$^{53}$] Osservatorio Astrofisico di Torino (INAF), Torino, Italy
\item[$^{54}$] Politecnico di Milano, Dipartimento di Scienze e Tecnologie Aerospaziali , Milano, Italy
\item[$^{55}$] Universit\`a del Salento, Dipartimento di Matematica e Fisica ``E.\ De Giorgi'', Lecce, Italy
\item[$^{56}$] Universit\`a dell'Aquila, Dipartimento di Scienze Fisiche e Chimiche, L'Aquila, Italy
\item[$^{57}$] Universit\`a di Catania, Dipartimento di Fisica e Astronomia ``Ettore Majorana``, Catania, Italy
\item[$^{58}$] Universit\`a di Milano, Dipartimento di Fisica, Milano, Italy
\item[$^{59}$] Universit\`a di Napoli ``Federico II'', Dipartimento di Fisica ``Ettore Pancini'', Napoli, Italy
\item[$^{60}$] Universit\`a di Palermo, Dipartimento di Fisica e Chimica ''E.\ Segr\`e'', Palermo, Italy
\item[$^{61}$] Universit\`a di Roma ``Tor Vergata'', Dipartimento di Fisica, Roma, Italy
\item[$^{62}$] Universit\`a Torino, Dipartimento di Fisica, Torino, Italy
\item[$^{63}$] Benem\'erita Universidad Aut\'onoma de Puebla, Puebla, M\'exico
\item[$^{64}$] Unidad Profesional Interdisciplinaria en Ingenier\'\i{}a y Tecnolog\'\i{}as Avanzadas del Instituto Polit\'ecnico Nacional (UPIITA-IPN), M\'exico, D.F., M\'exico
\item[$^{65}$] Universidad Aut\'onoma de Chiapas, Tuxtla Guti\'errez, Chiapas, M\'exico
\item[$^{66}$] Universidad Michoacana de San Nicol\'as de Hidalgo, Morelia, Michoac\'an, M\'exico
\item[$^{67}$] Universidad Nacional Aut\'onoma de M\'exico, M\'exico, D.F., M\'exico
\item[$^{68}$] Institute of Nuclear Physics PAN, Krakow, Poland
\item[$^{69}$] University of \L{}\'od\'z, Faculty of High-Energy Astrophysics,\L{}\'od\'z, Poland
\item[$^{70}$] Laborat\'orio de Instrumenta\c{c}\~ao e F\'\i{}sica Experimental de Part\'\i{}culas -- LIP and Instituto Superior T\'ecnico -- IST, Universidade de Lisboa -- UL, Lisboa, Portugal
\item[$^{71}$] ``Horia Hulubei'' National Institute for Physics and Nuclear Engineering, Bucharest-Magurele, Romania
\item[$^{72}$] Institute of Space Science, Bucharest-Magurele, Romania
\item[$^{73}$] Center for Astrophysics and Cosmology (CAC), University of Nova Gorica, Nova Gorica, Slovenia
\item[$^{74}$] Experimental Particle Physics Department, J.\ Stefan Institute, Ljubljana, Slovenia
\item[$^{75}$] Universidad de Granada and C.A.F.P.E., Granada, Spain
\item[$^{76}$] Instituto Galego de F\'\i{}sica de Altas Enerx\'\i{}as (IGFAE), Universidade de Santiago de Compostela, Santiago de Compostela, Spain
\item[$^{77}$] IMAPP, Radboud University Nijmegen, Nijmegen, The Netherlands
\item[$^{78}$] Nationaal Instituut voor Kernfysica en Hoge Energie Fysica (NIKHEF), Science Park, Amsterdam, The Netherlands
\item[$^{79}$] Stichting Astronomisch Onderzoek in Nederland (ASTRON), Dwingeloo, The Netherlands
\item[$^{80}$] Universiteit van Amsterdam, Faculty of Science, Amsterdam, The Netherlands
\item[$^{81}$] Case Western Reserve University, Cleveland, OH, USA
\item[$^{82}$] Colorado School of Mines, Golden, CO, USA
\item[$^{83}$] Department of Physics and Astronomy, Lehman College, City University of New York, Bronx, NY, USA
\item[$^{84}$] Michigan Technological University, Houghton, MI, USA
\item[$^{85}$] New York University, New York, NY, USA
\item[$^{86}$] University of Chicago, Enrico Fermi Institute, Chicago, IL, USA
\item[$^{87}$] University of Delaware, Department of Physics and Astronomy, Bartol Research Institute, Newark, DE, USA
\item[] -----
\item[$^{a}$] Max-Planck-Institut f\"ur Radioastronomie, Bonn, Germany
\item[$^{b}$] also at Kapteyn Institute, University of Groningen, Groningen, The Netherlands
\item[$^{c}$] School of Physics and Astronomy, University of Leeds, Leeds, United Kingdom
\item[$^{d}$] Fermi National Accelerator Laboratory, Fermilab, Batavia, IL, USA
\item[$^{e}$] Pennsylvania State University, University Park, PA, USA
\item[$^{f}$] Colorado State University, Fort Collins, CO, USA
\item[$^{g}$] Louisiana State University, Baton Rouge, LA, USA
\item[$^{h}$] now at Graduate School of Science, Osaka Metropolitan University, Osaka, Japan
\item[$^{i}$] Institut universitaire de France (IUF), France
\item[$^{j}$] now at Technische Universit\"at Dortmund and Ruhr-Universit\"at Bochum, Dortmund and Bochum, Germany
\end{description}

\section*{Acknowledgments}

\begin{sloppypar}
The successful installation, commissioning, and operation of the Pierre
Auger Observatory would not have been possible without the strong
commitment and effort from the technical and administrative staff in
Malarg\"ue. We are very grateful to the following agencies and
organizations for financial support:
\end{sloppypar}

\begin{sloppypar}
Argentina -- Comisi\'on Nacional de Energ\'\i{}a At\'omica; Agencia Nacional de
Promoci\'on Cient\'\i{}fica y Tecnol\'ogica (ANPCyT); Consejo Nacional de
Investigaciones Cient\'\i{}ficas y T\'ecnicas (CONICET); Gobierno de la
Provincia de Mendoza; Municipalidad de Malarg\"ue; NDM Holdings and Valle
Las Le\~nas; in gratitude for their continuing cooperation over land
access; Australia -- the Australian Research Council; Belgium -- Fonds
de la Recherche Scientifique (FNRS); Research Foundation Flanders (FWO),
Marie Curie Action of the European Union Grant No.~101107047; Brazil --
Conselho Nacional de Desenvolvimento Cient\'\i{}fico e Tecnol\'ogico (CNPq);
Financiadora de Estudos e Projetos (FINEP); Funda\c{c}\~ao de Amparo \`a
Pesquisa do Estado de Rio de Janeiro (FAPERJ); S\~ao Paulo Research
Foundation (FAPESP) Grants No.~2019/10151-2, No.~2010/07359-6 and
No.~1999/05404-3; Minist\'erio da Ci\^encia, Tecnologia, Inova\c{c}\~oes e
Comunica\c{c}\~oes (MCTIC); Czech Republic -- GACR 24-13049S, CAS LQ100102401,
MEYS LM2023032, CZ.02.1.01/0.0/0.0/16{\textunderscore}013/0001402,
CZ.02.1.01/0.0/0.0/18{\textunderscore}046/0016010 and
CZ.02.1.01/0.0/0.0/17{\textunderscore}049/0008422 and CZ.02.01.01/00/22{\textunderscore}008/0004632;
France -- Centre de Calcul IN2P3/CNRS; Centre National de la Recherche
Scientifique (CNRS); Conseil R\'egional Ile-de-France; D\'epartement
Physique Nucl\'eaire et Corpusculaire (PNC-IN2P3/CNRS); D\'epartement
Sciences de l'Univers (SDU-INSU/CNRS); Institut Lagrange de Paris (ILP)
Grant No.~LABEX ANR-10-LABX-63 within the Investissements d'Avenir
Programme Grant No.~ANR-11-IDEX-0004-02; Germany -- Bundesministerium
f\"ur Bildung und Forschung (BMBF); Deutsche Forschungsgemeinschaft (DFG);
Finanzministerium Baden-W\"urttemberg; Helmholtz Alliance for
Astroparticle Physics (HAP); Helmholtz-Gemeinschaft Deutscher
Forschungszentren (HGF); Ministerium f\"ur Kultur und Wissenschaft des
Landes Nordrhein-Westfalen; Ministerium f\"ur Wissenschaft, Forschung und
Kunst des Landes Baden-W\"urttemberg; Italy -- Istituto Nazionale di
Fisica Nucleare (INFN); Istituto Nazionale di Astrofisica (INAF);
Ministero dell'Universit\`a e della Ricerca (MUR); CETEMPS Center of
Excellence; Ministero degli Affari Esteri (MAE), ICSC Centro Nazionale
di Ricerca in High Performance Computing, Big Data and Quantum
Computing, funded by European Union NextGenerationEU, reference code
CN{\textunderscore}00000013; M\'exico -- Consejo Nacional de Ciencia y Tecnolog\'\i{}a
(CONACYT) No.~167733; Universidad Nacional Aut\'onoma de M\'exico (UNAM);
PAPIIT DGAPA-UNAM; The Netherlands -- Ministry of Education, Culture and
Science; Netherlands Organisation for Scientific Research (NWO); Dutch
national e-infrastructure with the support of SURF Cooperative; Poland
-- Ministry of Education and Science, grants No.~DIR/WK/2018/11 and
2022/WK/12; National Science Centre, grants No.~2016/22/M/ST9/00198,
2016/23/B/ST9/01635, 2020/39/B/ST9/01398, and 2022/45/B/ST9/02163;
Portugal -- Portuguese national funds and FEDER funds within Programa
Operacional Factores de Competitividade through Funda\c{c}\~ao para a Ci\^encia
e a Tecnologia (COMPETE); Romania -- Ministry of Research, Innovation
and Digitization, CNCS-UEFISCDI, contract no.~30N/2023 under Romanian
National Core Program LAPLAS VII, grant no.~PN 23 21 01 02 and project
number PN-III-P1-1.1-TE-2021-0924/TE57/2022, within PNCDI III; Slovenia
-- Slovenian Research Agency, grants P1-0031, P1-0385, I0-0033, N1-0111;
Spain -- Ministerio de Ciencia e Innovaci\'on/Agencia Estatal de
Investigaci\'on (PID2019-105544GB-I00, PID2022-140510NB-I00 and
RYC2019-027017-I), Xunta de Galicia (CIGUS Network of Research Centers,
Consolidaci\'on 2021 GRC GI-2033, ED431C-2021/22 and ED431F-2022/15),
Junta de Andaluc\'\i{}a (SOMM17/6104/UGR and P18-FR-4314), and the European
Union (Marie Sklodowska-Curie 101065027 and ERDF); USA -- Department of
Energy, Contracts No.~DE-AC02-07CH11359, No.~DE-FR02-04ER41300,
No.~DE-FG02-99ER41107 and No.~DE-SC0011689; National Science Foundation,
Grant No.~0450696, and NSF-2013199; The Grainger Foundation; Marie
Curie-IRSES/EPLANET; European Particle Physics Latin American Network;
and UNESCO.
\end{sloppypar}

}

\end{document}